\documentstyle[12pt]{article}
\begin{document}
\vskip .7cm
\begin{center}
{\bf { DUAL BRST SYMMETRY FOR QED}}

\vskip 2cm

 R. P. MALIK
\footnote{ E-mail address: malik@boson.bose.res.in  }\\
{\it S. N. Bose National Centre for Basic Sciences,} \\
{\it Block-JD, Sector-III, Salt Lake, Calcutta-700 098, India} \\

\vskip 2.5cm

\end{center}

\noindent
{\bf Abstract:}               
We show the existence of a co(dual)-BRST symmetry for the usual BRST
invariant Lagrangian density of an Abelian  gauge theory in two dimensions
of space-time where a $U(1)$ gauge field is coupled to the Noether
conserved current (constructed by the Dirac fields). Under this new symmetry,
it is the gauge-fixing term that remains invariant and the symmetry
transformations on the Dirac fields are analogous to the chiral transformations.
This interacting theory is shown to provide a tractable field theoretical model for the Hodge theory.  The Hodge dual operation is shown to correspond to 
a discrete symmetry in the theory and the extended BRST algebra for the 
generators of the underlying symmetries turns out to be reminiscent of the 
algebra obeyed by the de Rham cohomology operators of differential geometry.

\baselineskip=16pt

\vskip 1cm

\newpage

\noindent
{\bf 1 Introduction}\\

\noindent
The study of symmetries in theoretical physics has brought together many 
physical ideas and underlying mathematical concepts on an equal footing where 
both have enriched each-other in a grand and illuminating manner. One such 
symmetry is the  Becchi-Rouet-Stora-Tyutin (BRST) symmetry [1,2] which has 
played a key role in bringing together the physics of gauge theory (endowed 
with the first-class constraints) and mathematics of differential geometry 
(related with connections and curvature tensors, etc.) in the 
realm of what is known as the BRST cohomology. In fact,
the nilpotent ($Q_{B}^2 = 0$) BRST charge 
($Q_{B}$) is analogous to the exterior derivative $d$ ($ d^2 = 0$)
of differential geometry and BRST closed states ($ Q_{B} |phys> = 0$) are
the {\it physical states} which are analogous to the
closed-forms $f$ ($ d f = 0$) of differential geometry. The subsidiary condition
$Q_{B} |phys> = 0$ is required to ensure that the 
operator form of first-class constraints 
of the original gauge theories must annihilate the physical states of the
theory (Dirac's prescription) [3-8]. Now two physical states are said to
belong to the same cohomology class with respect to BRST charge $Q_{B}$ if
they differ by a BRST exact state. There are two other operators of
differential geometry that are required in the definition of the Hodge
decomposition theorem which states that, on a compact manifold, 
any $n$-form $f_{n} (n = 0, 1, 2...)$
can be uniquely written as the sum of a harmonic form $h_{n}$
($\Delta h_{n} = 0, d h_{n} = 0, \delta h_{n} = 0$), an exact form $d e_{n -1}$
and a co-exact form $\delta c_{n +1}$ as
$$
\begin{array}{lcl}
f_{n} = h_{n} + d\; e_{n-1} + \delta \;c_{n+1},
\end{array}\eqno (1.1)
$$
where $\delta$ ($ \delta = \pm * d *; \delta^2 = 0$) is the co(dual)-exterior
derivative and $\Delta$ ($ \Delta = (d + \delta)^2 = d \delta + \delta d$) is
the Laplacian operator [7-13]. It has been a long-standing problem to express
$\delta$ and $\Delta$ in the language of some {\it local} symmetry properties 
of a given
BRST invariant Lagrangian density for a gauge theory. Some very interesting and
illuminating attempts have been made towards this goal for the {\it interacting}
(non)Abelian gauge theories in any arbitrary dimension of space-time
but symmetry transformations have turned out to be non-local and non-covariant
[14-17]. In the covariant formulation [18], the nilpotency is restored only
for the specific values of the parameters of the theory.

Recently, it has been shown [19,20] that free two-dimensional (2D) Abelian-
and self-interacting non-Abelian gauge theories (without any interaction with 
matter fields) 
are the prototype examples of a new kind of topological field theories
\footnote{A theory with a flat space-time metric but without any propagating
degrees of freedom. It captures together some of the key features of Witten- 
and Schwarz-type of topological field theories.} [21] that provide a field 
theoretical model for the Hodge theory where all the de Rham cohomology 
operators ($d , \delta, \Delta$) correspond to
the {\it generators} of local, continuous, covariant and nilpotent (for $d$ and 
$\delta$) symmetries of the theory. Such an analogy has also been established
for the free two-form Abelian gauge theory in four ($ 3 + 1$) dimensional
space-time [22]. All these above examples of free gauge theories are, however, 
ideal as well as simple examples and have very little to do with physical
reality. The most natural and physically interesting gauge theories are the 
interacting 
theories. It is, therefore, a nice idea to explore the possibility of having an
interacting field theory as a tractable model for the Hodge theory. The 
purpose of the present paper is to show that 2D quantum electrodynamics 
(QED) (a dynamically closed system of $U(1)$ gauge field $A_{\mu}$ and Dirac 
fields $\psi$ and $\bar \psi$) is an {\it interacting}
field theoretical model for the Hodge theory. Besides the local BRST charge
(analogue of $d$), we show that there are local co-BRST charge $Q_{D}$ 
(analogue of $\delta$) and a local bosonic (Casimir) operator $W$ (analogue of
$\Delta$) that are generators of symmetry transformations for the BRST
invariant Lagrangian density of QED. Like two-form $ F = d A$ remains invariant
under the BRST transformations and the Dirac fields transform to the analogue 
of the local gauge transformations, it is the gauge-fixing term  
\footnote{The vector potential $A_{\mu}$ of the $U(1)$ gauge theory
is defined through one-form $A= A_{\mu} dx^{\mu}$. The gauge-fixing
term $ \partial \cdot A = \delta A $ is Hodge dual to the two-form $F 
= d A $ where $ \delta = \pm * d * $ is the adjoint(dual) 
exterior derivative and $d$ is the exterior derivative (see, e.g.,
Refs. [9,10]).} that remains invariant under the dual BRST transformations 
and the Dirac fields transform as an analogue of the chiral transformations.
We show that the existence of a discrete symmetry connects these two 
symmetry transformations and turns out to be the analogue of the Hodge dual
$*$ operation of the differential geometry. Finally, we show that the extended
BRST algebra constituted by generators of all the underlying symmetries of
the theory mimic the algebra of the de Rham cohomology operators of 
differential geometry. To the best of our knowledge, this {\it interacting}
field theory is the first example of a Hodge theory where all the de Rham
cohomology operators correspond to some {\it local} generators.

The outline of this paper is as follows. In Sec. 2, we recapitulate the bare
essentials of the BRST symmetries and set up the notations. This is followed
by the discussions of dual BRST symmetries in Sec. 3. Section 4 is devoted
to the discussion of symmetry generated by the Casimir operator. 
In Sec. 5, we discuss a discrete symmetry that corresponds to the Hodge duality
operation. The extended BRST algebra and the Hodge decomposition theorem 
are discussed in Sec. 6. Finally, we make some concluding remarks in Sec. 7.\\

\noindent
{\bf 2 BRST Symmetries}\\

\noindent
Let us begin with a $D$-dimensional BRST invariant Lagrangian density 
(${\cal L}_{B}$) for
the interacting $U(1)$ gauge theory in the Feynman gauge:
$$
\begin{array}{lcl}
{\cal L}_{B} = - \frac{1}{4} F^{\mu\nu} F_{\mu\nu}
+ \bar \psi (i \gamma^{\mu} \partial_{\mu} - m) \psi
- e \bar \psi \gamma^{\mu} A_{\mu} \psi
+ B (\partial \cdot A) + \frac{1}{2} B^2 - i \partial_{\mu} \bar C
\partial^{\mu} C,
\end{array}\eqno (2.1)
$$
where $F_{\mu\nu} = \partial_{\mu} A_{\nu} - \partial_{\nu} A_{\mu}$
is the field strength tensor, $B$ is the Nakanishi-Lautrup auxiliary field,
$(\bar C)C$ are the anticommuting Faddeev-Popov (anti)ghost fields ($ \bar C^2 
= C^2 = 0, C \psi = - \psi C, \bar C \psi = - \psi \bar C, C \bar C
= - \bar C C, etc.$) and
indices $\mu,\nu = 0, 1, 2,......D-1$ represent the flat Minkowski space-time
directions. It can be checked that the above Lagrangian density remains
quasi-invariant ($\delta_{B} {\cal L}_{B} = \eta \partial_{\mu}
[B \partial^{\mu} C]$) under the following off-shell nilpotent
($ \delta_{B}^2 = 0$) BRST transformations:    
$$
\begin{array}{lcl}
\delta_{B} A_{\mu} &=& \eta\; \partial_{\mu}\; C, \;\;\qquad \;\;
\delta_{B} C = 0, \; \;\qquad \; \;\delta_{B} \bar C 
= i \;\eta\; B, \;\;\qquad \;\;
\delta_{B} B = 0, \nonumber\\
\delta_{B} \psi &=& - i \eta e C \psi, \quad\;
\delta_{B} \bar \psi = i \eta e C \bar \psi, \quad \;
\delta_{B} (\partial \cdot A) = \eta \Box C, \quad \;
\delta_{B} F_{\mu\nu} = 0,
\end{array}\eqno (2.2)
$$
where $\eta$ is an anticommuting ($ \eta C = - C \eta, \eta \bar C =
- \bar C \eta, \eta \psi = - \psi \eta, \eta \bar \psi = - \bar \psi \eta$)
space-time independent transformation parameter. The on-shell 
($ \Box C = 0$) nilpotent ($\delta_{b}^2 = 0$) BRST transformations can be
obtained from (2.1) by inserting the equations of motion for the $B$ field
( i.e. $ B = - ( \partial \cdot A)$ ) as
$$
\begin{array}{lcl}
\delta_{b} A_{\mu} &=& \eta\; \partial_{\mu}\; C, \;\;\qquad \;\;
\delta_{b} C = 0, \; \;\qquad \; \;\delta_{b} \bar C 
= - i \;\eta\; (\partial \cdot A), \;\;\qquad \;\; \nonumber\\
\delta_{b} \psi &=& - i \eta e C \psi, \quad\;
\delta_{b} \bar \psi = i \eta e C \bar \psi, \quad \;
\delta_{b} (\partial \cdot A) = \eta \Box C, 
\end{array}\eqno (2.3)
$$
and the corresponding on-shell nilpotent BRST invariant Lagrangian
density is
$$
\begin{array}{lcl}
{\cal L}_{b} = - \frac{1}{4} F^{\mu\nu} F_{\mu\nu}
+ \bar \psi (i \gamma^{\mu} \partial_{\mu} - m) \psi
- e \bar \psi \gamma^{\mu} A_{\mu} \psi
- \frac{1}{2} (\partial \cdot A)^2 - i \partial_{\mu} \bar C
\partial^{\mu} C.
\end{array}\eqno (2.4)
$$
It is straightforward to check that the generator for the above symmetry
transformations is the off(on)-shell nilpotent ($Q_{B,(b)}^2 = 0$) BRST charge
$$
\begin{array}{lcl}
 Q_{B} &=& {\displaystyle \int} d^{(D-1)} x
\bigl [ \partial_{i} F^{0i} C + B \dot C 
- e \bar \psi \gamma_{0} C \psi \bigr ] 
\equiv {\displaystyle \int} d^{(D-1)} x \bigl [ B \dot C - \dot B C \bigr ],
\nonumber\\ 
 Q_{b} &=& {\displaystyle \int} d^{(D-1)} x
\bigl [ \partial_{i} F^{0i} C - (\partial \cdot A) \dot C 
- e \bar \psi \gamma_{0} C \psi \bigr ] \nonumber\\
&\equiv& {\displaystyle \int} d^{(D-1)} x \bigl [\partial_{0}
(\partial \cdot A) C - (\partial \cdot A) \dot C \bigr ],
\end{array} \eqno(2.5)
$$
where the latter expressions for $Q_{B,b}$ have been obtained by exploiting
the equation of motions: ($ \partial_{\mu} F^{\mu\nu} - \partial^{\nu} B
= e \bar \psi \gamma^{\nu} \psi $) and
($ \partial_{\mu} F^{\mu\nu} + \partial^{\nu} (\partial \cdot A)
= e \bar \psi \gamma^{\nu} \psi $). The global scale invariance of (2.1)
under $ C \rightarrow e^{- \lambda} C, \bar C \rightarrow e^{\lambda}
\bar C, A_{\mu} \rightarrow A_{\mu}, B \rightarrow B $ (where $\lambda$
is a global parameter), leads to the derivation of a conserved ghost charge
($Q_{g}$)
$$
\begin{array}{lcl}
Q_{g} 
= - i {\displaystyle \int} d^{(D-1)} x \;\bigl [\; C \dot {\bar C} 
+ \bar C \dot  C
\; \bigr ].
\end{array} \eqno(2.6)
$$
Together, these conserved charges satisfy the following algebra
$$
\begin{array}{lcl}
Q_{B}^2 &=& \frac{1}{2} \{ Q_{B}, Q_{B} \} = 0, \quad
i [ Q_{g}, Q_{AB} ] = - Q_{AB}, \quad
i [ Q_{g}, Q_{B} ] = + Q_{B}, \nonumber\\
Q_{AB}^2 &=& \frac{1}{2} \{ Q_{AB}, Q_{AB} \} = 0, \qquad
\{ Q_{B}, Q_{AB} \}  = 0,
\end{array} \eqno(2.7)
$$
where $Q_{AB}$ is the anti-BRST charge which can be readily obtained from (2.5)
by the replacement :$ C \rightarrow i \bar C$. Note that
$ C \rightarrow \pm i \bar C, \bar C \rightarrow \pm i C$ is the discrete
symmetry of the ghost action ($ I_{F.P.} = - i \int d^{D} x\;
\partial_{\mu} \bar C \partial^{\mu} C$) in any arbitrary dimension of 
spacetime. This is why the anti-BRST charge $Q_{AB}$ and the corresponding
nilpotent transformations can be obtained  from (2.5) and (2.2) by the
replacement: $ C \rightarrow \pm i \bar C$.

At this stage, it is worth pointing out the fact that the de Rham cohomology
operators ($ d , \delta, \Delta$) obey the following algebra
$$
\begin{array}{lcl}
&& d^2 = 0, \qquad \delta^2 = 0, \qquad \Delta = (d + \delta)^2 = d \delta
+ \delta d, \nonumber\\
&& [ \Delta, d ] = 0, \qquad [ \Delta , \delta ] = 0 \qquad
 \{ d, \delta \} = \Delta \neq 0.
\end{array} \eqno(2.8)
$$
It can be easily seen that if $Q_{B}$ is the analogue of $d$, $Q_{AB}$ is 
{\it not} the analogue of $\delta$ because $Q_{B}$ and $Q_{AB}$ anticommute 
\footnote{Cohomologically higher-order BRST- and anti-BRST operators do not
anticommute for the compact non-Abelian Lie algebras [23]. Our discussion,
however, concerns with a simple Abelian Lie algebra.}
with each-other whereas $d$ and $\delta$ do {\it not} anticommute. Thus, in
the algebra (2.7), there is no analogue of the Laplacian operator $\Delta$.
Furthermore, under transformations generated by $Q_{B}$ and $Q_{AB}$, it
is the two-form $ F = d A$ (derived from the application of $d$
on the one-form $A$) that remains invariant and not the gauge-fixing term
(which is derived from one-form $A$ by application of dual exterior derivative
$\delta$). Thus, it is obvious that, at the algebraic- and conceptual level, 
the analogy is not complete between the BRST algebra (2.7) and the de Rham 
cohomology algebra (2.8).\\

\noindent
{\bf 3 Dual BRST symmetries}\\

\noindent
In two ($ 1 + 1$) dimensions of space-time, there exists only one component
(i.e. electric field $E = F_{01}$) of the field strength tensor $F_{\mu\nu}$
and there is no magnetic field.
Thus, the analogue of the BRST invariant Lagrangian density (2.1) is:
$$
\begin{array}{lcl}
{\cal L}_{B} =  \frac{1}{2} \; E^2
+ \bar \psi (i \gamma^{\mu} \partial_{\mu} - m) \psi
- e \bar \psi \gamma^{\mu} A_{\mu} \psi
+ B (\partial \cdot A) + \frac{1}{2} B^2 - i \partial_{\mu} \bar C
\partial^{\mu} C,
\end{array}\eqno (3.1a)
$$
which can be recast as:
$$
\begin{array}{lcl}
{\cal L_{B}} =  {\cal B} E - \frac{1}{2} \; {\cal B}^2
+ \bar \psi (i \gamma^{\mu} \partial_{\mu} - m) \psi
- e \bar \psi \gamma^{\mu} A_{\mu} \psi
+ B (\partial \cdot A) + \frac{1}{2} B^2 - i \partial_{\mu} \bar C
\partial^{\mu} C,
\end{array}\eqno (3.1b)
$$
by introducing another auxiliary field ${\cal B}$. It can be checked that
under the following off-shell nilpotent ($\delta_{D}^2 = 0$) dual-BRST 
transformations
$$
\begin{array}{lcl}
\delta_{D} A_{\mu} &=& -\eta \varepsilon_{\mu\nu}
\partial^{\nu} \bar C, \quad\;
\delta_{D} \bar C = 0, \quad \;\delta_{D}  C = - i \eta {\cal B}, \quad\;
\delta_{D} {\cal B} = 0, \quad \;\delta_{D} B = 0, \nonumber\\
\delta_{D} \psi &=& - i \eta e \bar C \gamma_{5} \psi, \quad\;
\delta_{D} \bar \psi = i \eta e \bar C \gamma_{5} \bar \psi, \quad\;
\delta_{D} (\partial \cdot A) = 0, \quad\;
\delta_{D} E = \eta \Box {\bar C}, 
\end{array}\eqno (3.2)
$$
the Lagrangian density (3.1b) (with $m = 0$) transforms as:
$\delta_{D} {\cal L_{B}} = \eta \partial_{\mu} ({\cal B} \partial^{\mu}
\bar C)$
\footnote{ We adopt the notations in which the flat 2D Minkowski metric
$\eta_{\mu\nu} = $ diag$\; (+1, -1), \gamma^{0} = \sigma_{2}, 
\gamma^{1} = i \sigma_{1}, \gamma_{5} = \gamma^{0} \gamma^{1} = \sigma_{3},
\{ \gamma^{\mu}, \gamma^{\nu} \} = 2 \eta^{\mu\nu}, \gamma_{\mu} \gamma_{5}
= \varepsilon_{\mu\nu} \gamma^{\nu}, \varepsilon_{01} = \varepsilon^{10}
= + 1, F_{01} = \partial_{0} A_{1} - \partial_{1} A_{0} 
= E = - \varepsilon^{\mu\nu} \partial_{\mu} A_{\nu} = F^{10},
\Box = \eta^{\mu\nu} \partial_{\mu} \partial_{\nu} = (\partial_{0})^2
- (\partial_{1})^2 $ and  here $\sigma's$ are the usual
$ 2 \times 2$ Pauli matrices.}. The on-shell ($ \Box \bar C = 0$)
nilpotent ($ \delta_{d}^2 = 0$) dual-BRST symmetry transformations (for $ m 
= 0$) can be obtained from (3.2) by exploiting equations of motions w.r.t. $B$ 
and ${\cal B}$ fields (i.e. $ B = - (\partial \cdot A); {\cal B } = E$) as
$$
\begin{array}{lcl}
\delta_{d} A_{\mu} &=& -\eta \varepsilon_{\mu\nu}
\partial^{\nu} \bar C, \quad\;
\delta_{d} \bar C = 0, \quad \;\delta_{d}  C = - i \eta E, 
\quad \; \delta_{d} E = \eta \Box {\bar C}, \nonumber\\
\delta_{d} \psi &=& - i \eta e \bar C \gamma_{5} \psi, \quad\;
\delta_{d} \bar \psi = i \eta e \bar C \gamma_{5} \bar \psi, \quad\;
\delta_{d} (\partial \cdot A) = 0. 
\end{array}\eqno (3.3)
$$
The above on-shell ($ \Box \bar C = 0$) nilpotent transformations turn out 
to be the symmetry for the analogue of the Lagrangian density (2.4) 
$$
\begin{array}{lcl}
{\cal L}_{b} =  \frac{1}{2} \; E^2
+ \bar \psi (i \gamma^{\mu} \partial_{\mu} - m) \psi
- e \bar \psi \gamma^{\mu} A_{\mu} \psi
- \frac{1}{2} (\partial \cdot A)^2 - i \partial_{\mu} \bar C
\partial^{\mu} C.
\end{array}\eqno (3.4)
$$
We christen the above off(on)-shell nilpotent symmetries (3.2) and (3.3) as
dual-BRST symmetries because it is the gauge-fixing term that remains invariant
under these transformations and the corresponding symmetry transformations on
the Dirac fields are the analogue of the chiral transformations. It should be
contrasted with the usual BRST symmetries, under which, it is the electric 
field $E$ that remains invariant and the corresponding symmetry transformations on the Dirac fields are the analogue of the local gauge transformations
\footnote{As explained in Sec. 1, in two dimensional space-time, the 
electric field $E$ and gauge-fixing term $(\partial \cdot A)$ are `Hodge dual' 
to each-other. The gauge symmetry and chiral symmetry transformations on the 
Dirac fields are also dual to each-other (see, e.g., Ref. [24]).}.

It can be checked that the above transformations are generated by
$$
\begin{array}{lcl}
Q_{D} &=& {\displaystyle \int} d x
\bigl [{\cal B} \dot {\bar C} + e \bar \psi \gamma_{1} \bar C \psi
- (\partial_{1} B) \bar C
\bigr ] \equiv {\displaystyle \int} d x \;\bigl [ \;{\cal B} \dot {\bar C} 
- \dot {\cal B} \bar C
\bigr ],\nonumber\\
Q_{d} &=& {\displaystyle \int} d x
\bigl [ E \dot {\bar C} + e \bar \psi \gamma_{1} \bar C \psi
+ (\partial_{1} (\partial \cdot A) ) \bar C
\bigr ] \equiv {\displaystyle \int} d x \;\bigl [ \;E \dot {\bar C} 
- \dot E \bar C \bigr ],
\end{array} \eqno(3.5)
$$
where the latter expressions for $Q_{D, d}$ have been obtained by using
the equation of motion ($ \varepsilon^{\mu\nu} \partial_{\nu}
{\cal B} + \partial^{\mu} B = - e \bar \psi \gamma^{\mu} \psi$) and
($ \varepsilon^{\mu\nu} \partial_{\nu}
E - \partial^{\mu} (\partial \cdot A) = - e \bar \psi \gamma^{\mu} \psi$)
for the photon field present in the Lagrangian density (3.1b) and (3.4). 
Using the following BRST quantization conditions (with $\hbar = c = 1$):
$$
\begin{array}{lcl}
&& [ A_{0} (x,t), B (y, t) ] = i \delta (x - y), 
 [ A_{1} (x, t), {\cal B} (y, t) ] = i \delta (x - y), \nonumber\\
&& \{ \psi (x, t), \psi^{\dagger} (y, t) \} = - \delta (x - y), 
 \{ C (x ,t), \dot {\bar C} (y, t) \} = \delta (x - y), \nonumber\\
&& \{ \bar C (x, t), \dot C(y, t) \} = - \delta (x -y), 
\end{array} \eqno(3.6)
$$
(and rest of the (anti)commutators are zero), it can be seen that $Q_{D}$ is
indeed the generator for the transformations (3.2)
\footnote{ Here, and in what follows, we shall concentrate only on the
off-shell nilpotent symmetries and corresponding generators. The case of 
on-shell nilpotent symmetries 
can be easily derived from here.} if we exploit the following relationship 
$$
\begin{array}{lcl}
\delta_{D} \Phi = - i \eta [ \Phi, Q_{D} ]_{\pm},
\end{array}\eqno(3.7)
$$
where $[\;,\;]_{\pm}$ stands for (anti)commutator for the generic field
$\Phi$ being (fermionic)bosonic in nature. It is straightforward to check
that $ Q_{D}^2 = \frac{1}{2} \{ Q_{D}, Q_{D} \} = 0$ due to (3.6). A simpler
way to see this fact is: $ \delta_{D} Q_{D} = - i \eta \{ Q_{D}, Q_{D} \} = 0$
by exploiting (3.2) and (3.5).\\

\noindent
{\bf 4 Symmetry generated by the Casimir operator}\\

\noindent
It is very natural to expect that the anticommutator of these two
transformations ($\{ \delta_{B}, \delta_{D} \} = \delta_{W}$) would
also be the symmetry transformation ($\delta_{W}$) for the Lagrangian
density (3.1b) (with $ m = 0$). This is indeed the case as can be seen
that under the following bosonic ($ \kappa = - i \eta \eta^{\prime}$)
transformations corresponding to $\delta_{W}$
$$
\begin{array}{lcl}
\delta_{W} A_{\mu} &=& \kappa (\partial_{\mu} {\cal B}
+ \varepsilon_{\mu\nu} \partial^{\nu} B), \qquad\;
\delta_{W} {\cal B} = 0, \qquad \;\delta_{W} B = 0,
\nonumber\\
\delta_{W} (\partial \cdot A) &=& \kappa \Box {\cal B}, \;\qquad
\delta_{W} E = - \kappa \Box B, \;\quad 
\delta_{W} C = 0, \;\quad \delta_{W} \bar C = 0,
\nonumber\\
\delta_{W} \psi &=& \kappa i e (\gamma_{5} B - {\cal B}) \psi, \;\qquad\;
\delta_{W} \bar \psi = - \kappa i e (\gamma_{5} B - {\cal B}) \bar \psi,
\end{array}\eqno(4.1)
$$
the Lagrangian density (3.1b) (with $ m = 0$) transforms as:
$ \delta_{W} {\cal L_{B}} = \kappa \partial_{\mu} [ B \partial^{\mu}
{\cal B} - {\cal B} \partial^{\mu} B ]$. Here $\eta$ and
$\eta^{\prime}$ are the transformation parameters corresponding to
the transformations $\delta_{B}$ and $\delta_{D}$ respectively. 
The generator for the above transformations is:
$$
\begin{array}{lcl}
W &=& {\displaystyle \int} dx \;
\bigl [\; B (\partial_{1} B + e \bar \psi \gamma_{1} \psi)
- {\cal B} ( \partial_{1} {\cal B} - e \bar \psi \gamma_{0} \psi)\;
\bigr ] 
\equiv {\displaystyle \int} dx\; \bigl [\; B \dot {\cal B} -
\dot B {\cal B} \;\bigr ],
\end{array}\eqno(4.2)
$$
where the latter expression for $W$ has been obtained due to the use
of equation of motion: $ \varepsilon^{\mu\nu} \partial_{\nu}
{\cal B} + \partial^{\mu} B = - e \bar \psi \gamma^{\mu} \psi$. As we shall
see below (Sec. 6), this bosonic symmetry generator is the Casimir operator 
for the full extended BRST algebra which plays an important role in the
representation theory for this algebra [11-13].
There are other simpler ways to derive the expression for $W$. For
instance, it can be seen that anticommutator of $Q_{B}$ and $Q_{D}$
leads to the derivation of $W$ (i.e. $ \{Q_{B}, Q_{D} \} = W$) if we
exploit the basic (anti)commutators (3.6). Furthermore,
since $Q_{B}$ and $Q_{D}$ are the generators for transformations
(2.2) and (3.2) respectively, it can be seen that the following 
relationships
$$
\begin{array}{lcl}
\delta_{D} Q_{B} &=& - i \eta \{ Q_{B}, Q_{D} \} = - i \eta W,
\nonumber\\
\delta_{B} Q_{D} &=& - i \eta \{ Q_{D}, Q_{B} \} = - i \eta W,
\end{array}\eqno (4.3)
$$
lead to the definition and derivation of $W$. \\

\noindent
{\bf 5 Discrete symmetries}\\

\noindent
It can be readily seen that, in 2D, the ghost action ($ I_{F.P} 
= - i \int d^2 x\; \partial_{\mu} \bar C \partial^\mu C$) remains
invariant under
$$
\begin{array}{lcl}
C \rightarrow \pm i \bar C, \quad \bar C \rightarrow \pm i C, \quad
\partial_{\mu} \rightarrow \pm i \varepsilon_{\mu\nu} \partial^{\nu}.
\end{array}\eqno (5.1)
$$
This symmetry was exploited for the definition of anti-BRST and
dual/anti-dual BRST symmetries in the case of {\it free} Abelian
gauge theories [19, 20]. Furthermore, it was noticed that under the 
following separate and independent transformations
$$
\begin{array}{lcl}
C \rightarrow \pm i \bar C, \quad \bar C \rightarrow \pm i C \quad
\partial_{\mu} \rightarrow \pm i \varepsilon_{\mu\nu} \partial^{\nu}
\quad A_{\mu} \rightarrow A_{\mu},
\end{array}\eqno (5.2)
$$
$$
\begin{array}{lcl}
C \rightarrow \pm i \bar C, \quad \bar C \rightarrow \pm i C \quad
A_{\mu} \rightarrow \mp i \varepsilon_{\mu\nu} A^{\nu},
\end{array}\eqno (5.3)
$$
the free (without matter fields $\psi$ and $\bar \psi$) Lagrangian density
(2.4) remains invariant. The above transformations were found to be the 
analogue of Hodge duality ($*$) operations for the free theory [19,20]. It
was also shown that the (anti)dual BRST transformations can be obtained
from the usual (anti)BRST transformations by exploiting the above discrete
symmetries. Similarly, two sets of topological invariants of the theory
were shown to be connected to each-other by (5.2) or (5.3).

For the interacting theory, described by the Lagrangian densities
(3.1a, 3.1b) and (3.4), similar transformations exist. It can be checked
that, under the following transformations
$$
\begin{array}{lcl}
C &\rightarrow& \pm i \gamma_{5} \bar C, \quad  A_{0} \rightarrow 
\pm i \gamma_{5} A_{1}, \quad E \rightarrow \pm i \gamma_{5} (\partial
\cdot A), \quad \psi \rightarrow \psi, \nonumber\\
\bar C &\rightarrow& \pm i \gamma_{5} C, \quad A_{1} \rightarrow \pm i
\gamma_{5} A_{0}, \quad (\partial \cdot A) \rightarrow \pm i \gamma_{5} E,
\quad \bar \psi \rightarrow \bar \psi, \nonumber\\
{\cal B} &\rightarrow& \mp i \gamma_{5} B, \qquad 
B \rightarrow \mp i \gamma_{5} {\cal B}, \qquad  e \rightarrow \mp i e,
\end{array}\eqno (5.4)
$$
the above Lagrangian densities (for the interacting theory) remain invariant
\footnote{Note that in the matrix notations, the transformations like:
$ A_{0} \rightarrow \pm i \gamma_{5} A_{1}$, etc., imply that $ A_{0}
\rightarrow \pm i A_{1}$ or/and $ A_{0} \rightarrow \mp i A_{1}$. Except for 
the exchange of signs, these transformations are same. In fact, $\gamma_{5}$ is
present appropriately in the rest of the transformations to take care of
this sign-flip.}.
Furthermore, it can be easily seen that the dual-BRST symmetries (3.2) and
(3.3) can be derived from their counterpart BRST symmetries ((2.2) and
(2.3)) by exploiting the symmetry transformations (5.4). Thus, as argued
in Refs. [19, 22], the discrete transformations in (5.4) are the analogue
of the Hodge duality ($*$) operation of differential geometry. To verify
this claim, we first note that, under two successive $*$ operations 
(corresponding to transformations (5.4)), the generic fields $\Phi$ of 
the above Lagrangian densities, transform as
$$
\begin{array}{lcl}
* \;\bigl (\; *\; \Phi \;\bigr ) = \pm \; \Phi,
\end{array}\eqno (5.5)
$$
where $(+)$ sign stands for the generic field $\Phi$ being $ \psi, \bar \psi$
and $(-)$ sign is for the rest of the fields (e.g.
$ \Phi = C, \bar C, B, {\cal B}, A_{0}, A_{1},
(\partial \cdot A), E$). Now, it is very gratifying to note that, for the
generic field $\Phi$, we have the following relationship
$$
\begin{array}{lcl}
\delta_{D}  \Phi = \;\pm \;* \;\delta_{B} \;*\; \Phi,
\end{array}\eqno (5.6)
$$
where $\delta_{D}$ and $\delta_{B}$ are the nilpotent symmetry 
transformations (3.2) and (2.2) respectively and $(+)$ sign stands for
$\Phi = \psi, \bar \psi$  and $(-)$ sign is for the rest of the fields.
The relative signs in (5.6) are dictated by the corresponding
signs in (5.5). Thus, we see that the relationship between transformations 
$\delta_{B}$ and $\delta_{D}$ acting 
on the generic field $\Phi$ is same as
the relationship between exterior derivative $d$ and the dual-exterior 
derivative $\delta$ (i.e. $\delta = \pm * d * $) acting on a 
differential form (defined on a compact manifold).

It is worthwhile to mention that under discrete transformations (5.4),
the generators ($Q_{(B, AB)}, Q_{(D, AD)}, W$) for the continuous 
transformations undergo the following change
$$
\begin{array}{lcl}
Q_{(B, AB)} \rightarrow Q_{(D, AD)}, \qquad Q_{(D, AD)} 
\rightarrow Q_{(B, AB)} \qquad W \rightarrow W,
\end{array}\eqno (5.7)
$$
so that the extended BRST algebra (see, e.g., eqn. (6.1) below) 
remains intact as far as these operators are concerned. Here $Q_{AD}$ is the 
anti-dual BRST charge which can be readily derived from (3.5) by the 
replacement :$ \bar C \rightarrow \pm i C$. The above transformations 
(5.7) should be contrasted with the 
four $ (3 + 1)$ dimensional case of the free 2-form Abelian gauge theories
[22] where it has been shown that under the analogue of $*$ operation, the
generators transform as: $ Q_{B} \rightarrow Q_{D}, Q_{D} \rightarrow
- Q_{B}, W \rightarrow - W$ which turns out to mimic the duality
symmetry transformation of the free Maxwell equations where $ {\bf E}
\rightarrow {\bf B}, {\bf B} \rightarrow - {\bf E}$. As expected by the
duality considerations in 2D [25-27], the Hodge duality $(*)$ transformations
in (5.7) are correct and consistent because of the fact that dualities in
4D and 2D are of different nature. In fact, it is precisely because of this
reason that, in 2D,  one obtains (a reciprocal relationship between
$\delta_{B}$ and $\delta_{D}$):
$$
\begin{array}{lcl}
\delta_{B}  \Phi = \;\pm \;* \;\delta_{D} \;*\; \Phi,
\end{array}\eqno (5.8)
$$
analogous to (5.6). This is a reflection of the fact that in 2D,
there is no relative change of sign between an operator and its dual
operator under the duality transformation.\\

\noindent
{\bf 6 Extended BRST algebra}\\

\noindent
Together, all the above
generators obey the following extended BRST algebra
$$
\begin{array}{lcl}
&& [ W, Q_{k} ] = 0, \;\;\;\qquad\;\;\; k = g, B, D, AB, AD, \nonumber\\
&& Q_{B}^2 = Q_{D}^2 = Q_{AB}^2 = Q_{AD}^2 = 0, \qquad
\{ Q_{D}, Q_{AD} \} = 0, \nonumber\\
&& \{ Q_{B}, Q_{D} \} = \{Q_{AB}, Q_{AD} \} = W, \qquad
\{ Q_{B}, Q_{AB} \} = 0, \nonumber\\
&& i [ Q_{g}, Q_{B} ] = Q_{B}, \qquad i [ Q_{g}, Q_{AB} ] = - Q_{AB},
\nonumber\\
&& i [ Q_{g}, Q_{D} ] = - Q_{D}, \qquad i [ Q_{g}, Q_{AD} ] = Q_{AD},
\end{array}\eqno(6.1)
$$
and the rest of the (anti)commutators are zero. It is evident that the
operator $W$ is the Casimir operator for the whole algebra. The 
{\it mathematical aspects} of the representation theory for the above kind 
of BRST algebra
have been discussed in Refs. [11-13]. It will be noticed that the ghost
number for $Q_{B}$ and $Q_{AD}$ is $+1$ and that of $Q_{D}$ and $Q_{AB}$
is $-1$. Now, given a state $| \phi> $  with ghost number $n$
 in the quantum Hilbert space (i.e.  $i Q_{g} | \phi > = n |\phi> $),
it can be readily seen, using the algebra (6.1), that
$$
\begin{array}{lcl}
i Q_{g} Q_{B} | \phi > &=& (n + 1) Q_{B} |\phi>, \qquad
i Q_{g} Q_{AD} | \phi > = ( n + 1) Q_{AD}|\phi>, \nonumber\\ 
i Q_{g} Q_{D} | \phi > &=& (n - 1) Q_{D} |\phi>, \qquad
i Q_{g} Q_{AB} | \phi > = (n - 1) Q_{AB} |\phi>, \nonumber\\
i Q_{g} W   \;| \phi > &=& n \;    W \;  |\phi>.
\end{array}\eqno(6.2)
$$   
This shows that the ghost numbers for the states $Q_{B} |\phi>$
(or $ Q_{AD} |\phi>$), $Q_{D} |\phi>$ (or $Q_{AB} |\phi>$) 
and $W |\phi>$ in the quantum Hilbert space are $ (n + 1), (n - 1)$ and $ n $ 
respectively.

At this stage, it is worth mentioning that in Refs. [19-21], the analogous
expressions for $Q_{B}, Q_{D}, W$ have been
derived for the free 2D Abelian- and non-Abelian gauge theories
(having no interaction with matter fields). The
topological properties of these free theories
have been shown to be encoded in the vanishing
of the operator $W$ when equations of motion are exploited. On the
contrary, as it turns out, the operator $W$ is defined
off-shell as well as on-shell for the 2D interacting BRST invariant
$U(1)$ gauge theory. This is because of the fact that even though
$U(1)$ gauge field is topological (i.e. without any propagating
degrees of freedom), it is coupled to the Dirac fields here
and fermionic degrees of freedom are present in the off-shell as well
as on-shell expression for $W$. Thus, the present theory is an example
of an {\it interacting} topological field theory in 2D.

It is interesting to note that the algebra of $Q_{B}, Q_{D}$ and $W$ in 
equation (6.1) is exactly identical to the corresponding algebra 
for the de Rham cohomology operators ($d, \delta, \Delta $) in (2.8).
Furthermore, it can be readily seen that 
the operation of these generators on a state with ghost number $n$
(cf. eqn.(6.2)) is same as the operation of the above cohomological
operators on a differential form of degree $n$. Thus, it 
is clear that  the BRST
cohomology can be defined comprehensively in terms of the above operators
and Hodge decomposition theorem
can be implemented cogently in the quantum Hilbert space of states where any
arbitrary state $ |\phi>_{n}$ (with ghost number $n$) 
can be written as the sum
of a harmonic state $|\omega>_{n}$ ($ W |\omega>_{n} = 0, Q_{B} |\omega>_{n}
= 0, Q_{D} |\omega>_{n} = 0$), a BRST exact state ($Q_{B} |\theta>_{n-1}$)
and a co-BRST exact state ($Q_{D} |\chi>_{n+1}$). 
Mathematically, this statement (which is the analogue of eqn. (1.1)) can be 
expressed by the following equation
$$
\begin{array}{lcl}
|\phi>_{n} = |\omega>_{n} +\; Q_{B} |\theta>_{n-1} + \;Q_{D} |\chi>_{n+1}.
\end{array}\eqno(6.3)
$$
It is obvious, therefore, that the above 
symmetry generators $Q_{B}, Q_{D}$ and $ W$ have
their counterparts in differential geometry as the de Rham Cohomology
operators $ d, \delta$ and $ \Delta$ respectively [9,10]. 
It is an interesting feature
of the BRST formalism that these de Rham cohomological operators can also be
identified with the generators $ Q_{AD}, Q_{AB}$ and $ W = \{ Q_{AD}, Q_{AB}
\} $ respectively. Thus, there is a two-to-one mapping from BRST formalism
to the differential geometry as: $(Q_{B}, Q_{AD}) \Rightarrow d,
(Q_{D}, Q_{AB}) \Rightarrow \delta, W = \{ Q_{B}, Q_{D} \} = \{ Q_{AD},
Q_{AB} \} \Rightarrow \Delta$. \\

\noindent
{\bf 7 Conclusions}\\

\noindent
We have shown that, in the language of well-defined symmetry properties of the 
BRST invariant Lagrangian density of an interacting $U(1)$ gauge theory, the 
celebrated Hodge decomposition theorem can be understood. The generators of
the nilpotent BRST transformations; nilpotent dual BRST transformations and 
a bosonic symmetry transformation (derived by taking the anticommutator of both these nilpotent symmetries) correspond to the de Rham cohomology operators
($ d, \delta, \Delta)$ of differential geometry and they obey the same kind
of algebra as obeyed by these operators. There is a logical 
understanding for the existence of these generators. In fact, these individual 
generators generate {\it separate and independent} symmetry transformations 
under which the two-form field (i.e. electric field $E$),
gauge-fixing term and ghost fields remain invariant respectively. A discrete
symmetry of the theory corresponds to the Hodge duality $(*)$ operation of the
differential geometry. Thus, all the cohomological quantities of differential
geometry find their analogue in the language of physically well-defined
quantities for interacting $U(1)$ gauge theory. Thus, this interacting theory
is a physical field theoretical model for the Hodge theory.
It will be very useful to explore the impact of this new symmetry
in the context of symmetries of the Green's functions for QED and derive
the analogue of Ward-Takahashi identities. Since the new symmetry is connected 
with the analogue of chiral transformation, it might be possible that BRST
cohomology and Hodge decomposition theorem will shed some light on the
Adler-Bardeen-Jackiw anomaly in 2D. It is quite possible that  this 
understanding might provide an insight into the  proof of the
consistency and unitarity of the anomalous gauge theory in two-dimensions
(see, e.g., Refs. [28,29] and references therein). The generalization of 
this new symmetry to 2D non-Abelian gauge theory (having local gauge 
interaction with matter fields) is another future direction that can be 
pursued. A very good attempt has been made to discuss these new symmetries
in the framework of Hamiltonian formulation [30]. However, more works are
needed in this direction to establish a precise analogy with the key concepts of
differential geometry.  The insights
gained in these studies might provide a clue for the generalization of
this new symmetry to physical four dimensional gauge theories. These are
some of the issues under investigation and a detailed discussion will be
reported elsewhere.\\

\noindent
{\bf Acknowledgments}\\

\noindent
A part of this work was done at JINR, Dubna (Moscow) and AS-ICTP, 
Trieste (Italy). Fruitful conversations with
members of theory groups at BLTP (JINR) and HEP group at
AS-ICTP are gratefully acknowledged.

\end{document}